\documentclass[]{raa}            % referee version: for submission
\usepackage{graphicx,times}
\usepackage{amssymb}
\usepackage{natbib}
\begin{document}%%
   \title{LAMOST J113208.06-005052.3 and LAMOST J052957.56+344127.0: two new binaries with a hot white dwarf and a flaring companion star}
\volnopage{ {\bf 0000} Vol.\ {\bf 0} No. {\bf XX}, 000--000}
\setcounter{page}{1}
\author{Y. H. Chen\inst{1,2,3}$^*$, \ C. M. Duan\inst{3,1}, \ and B. K. Sun\inst{1}}
\institute{\inst{1} Institute of Astrophysics, Chuxiong Normal University, Chuxiong 675000, China; {$yanhuichen1987@126.com$}\\
           \inst{2} International Centre of Supernovae (ICESUN), Yunnan Key Laboratory, Kunming 650216, China;\\
           \inst{3} Faculty of Science, Kunming University of Science and Technology, Kunming 650093, China\\
\vs \no
{\small Received [0000] [July] [day]; accepted [0000] [month] [day] }}

\abstract{Binaries contain rich physical information, and the study of binaries has always been a hot topic in stellar physics research. The stars LAMOST J1132 and LAMOST J0529 have not yet been recorded in the SIMBAD astronomical database. We have investigated their physical properties via methods such as spectral analysis, photometric analysis, and light curve analysis. Based on comprehensive analysis, we conclude that they are two newly discovered binary systems, each consisting of a hot white dwarf and a flaring companion star. Large Sky Area Multi-Object Fiber Spectroscopic Telescope (LAMOST) spectra indicate that both stars contain hot white dwarfs. The spectral fitting yields $T_{eff}$=53728$\pm$2467\,K, log$g$=7.98$\pm$0.08 for LAMOST J1132, and $T_{eff}$=47381$\pm$494\,K, log$g$=7.84$\pm$0.05 for LAMOST J0529. The weak neutral metal lines in the LAMOST spectra and the discrepancy between the Global Astrometric Interferometer for Astrophysics (GAIA) and LAMOST spectra both indicate that these two sources are likely binary systems. The relatively high flux values for both sources in the near-infrared and mid-infrared bands support our preliminary judgment. The color index in the near-infrared bands suggests that the companion star is K or M type for LAMOST J1132 and M type for LAMOST J0529. Light curve data from the Zwicky Transient Facility (ZTF) indicate that the companion stars of both sources are stars exhibiting flare activity. The eclipse probability is very low, indicating that these two sources are non-eclipsing binary systems. The physics of binaries is fascinating, and future data from LAMOST Medium Resolution Spectra are expected to enable the detection of magnetic fields in these two hot white dwarfs.
 }

\authorrunning{Y. H. Chen, C. M. Duan, \& B. K. Sun}       %author_head in even pages
\titlerunning{LAMOST J1132 and LAMOST J0529: Two New Binaries}  % title_head in odd pages
\maketitle

\section{Introduction}

When gazing at the starry sky at night, humans are filled with an infinite longing to explore the unknown, and the vast majority of celestial bodies they witness are stars. From a mass perspective, stars constitute only a small fraction of the total baryonic matter in the universe, with the majority being diffuse gas (Gupta et al. 2012). However, among all distinguishable, gravitationally bound independent celestial bodies in the universe (such as stars, planets, black holes, nebulae, etc.), stars account for the vast majority of the mass, making them the absolute dominant form of celestial bodies. White dwarfs (WDs) are the final stage of evolution for 98\% of all stars (Winget \& Kepler 2008), making their study of universal significance. Nuclear fusion has ceased in WDs, and their primary evolutionary process is cooling (which is accompanied by some contraction during the initial stages). WDs can even be over 12 billion years old (Torres et al. 2021), making them living fossils for studying galactic archaeology. WDs, composed of an electron-degenerate core and an ideal gas atmosphere, serve as natural laboratories for studying the laws of physics under extreme conditions. The precise asteroseismology of pulsating WDs (Giammichele et al. 2018) helps test and refine the theory of stellar structure and evolution. In summary, WDs encompass a wealth of physical principles, making their study profoundly significant.

In numerical simulation research, Machida et al. (2005) employed three-dimensional magnetohydrodynamic nested-grid simulations to detail the specific physical conditions under which rotating magnetized clouds fragment to form binaries. Two patterns of fragmentation are found, including ring fragmentation and bar fragmentation. Observational statistics show that binaries account for a significant proportion of stars. El-Badry \& Rix (2018) reported over 50,000 main-sequence (MS)/MS, more than 3,000 WD/MS, and nearly 400 WD/WD wide binaries within 200\,pc of the Sun from the Global Astrometric Interferometer for Astrophysics (GAIA, GAIA Collaboration et al. 2016) DR2. Spectral analysis is one of the effective methods for studying astrophysics. Qian et al. (2018) reported that as of June 16, 2017, about 3,196 EA-type binaries had been observed by the Large Sky Area Multi-Object Fiber Spectroscopic Telescope (LAMOST, Cui et al. 2012; Zhao et al. 2012), and their spectral types had been identified. Light curve analysis is uniquely advantageous for the study of variable sources. The Zwicky Transient Facility (ZTF, Bellm et al. 2019) is a telescope designed specifically for optical time-domain surveys using the Palomar 48 inch Schmidt telescope. Based on ZTF DR2 data, Chen et al. (2020) derived $\sim$350,000 eclipsing binaries. Research on binaries has always been a central focus in astrophysics.

In binaries, detached binaries consisting of a hot WD and a flaring companion star are relatively rare in observations. The reasons include the absence of ongoing mass transfer processes and the lack of significant photometric variability; their spectra are dominated by the hot WD, making the companion's spectral lines extremely faint even if detectable; and flares from the companion are intermittent, requiring long-term monitoring to be captured by chance. Nevertheless, such binaries are precursors to cataclysmic variables (CVs) and are critical for studying the detailed processes of binary star evolution. LAMOST J113208.06-005052.3 (hereafter LAMOST J1132) and LAMOST J052957.56+344127.0 (hereafter LAMOST J0529) were observed by LAMOST DR12 in its Low Resolution Search (LRS) on May 2, 2024, and October 18, 2023, respectively. These two sources were not observed again in either the LAMOST DR12 Medium Resolution Search (MRS) or the latest LAMOST DR13 LRS or MRS. We serendipitously discovered these two sources while studying magnetic WDs with spectra from the LAMOST survey. They have not been cataloged in the SIMBAD astronomical database. We have previously reported LAMOST J064137.77+045743.8 (Chen et al. 2025), which is also absent from the SIMBAD astronomical database, using a similar research methodology. Our comprehensive analysis reveals that both LAMOST J1132 and LAMOST J0529 are binaries, each harboring a hot WD and a companion star exhibiting flaring activity. They are observationally rare, rich in physics, and worthy of in-depth study. In Sect. 2, we perform a predictive study for LAMOST J1132 and LAMOST J0529 based on spectra from LAMOST and GAIA. A supplementary verification for the two stars based on photometric data are displayed in Sect. 3. In Sect. 4, we show a solid evidence of the companion star for LAMOST J1132 and LAMOST J0529 based on light curve data released by the ZTF telescope. In Sect. 5, we provide a thorough analysis, a concise discussion, and clear conclusions.

\section{A predictive study for LAMOST J1132 and LAMOST J0529 based on spectra from LAMOST and GAIA}

\begin{figure}
\begin{center}
\includegraphics[width=12cm,angle=0]{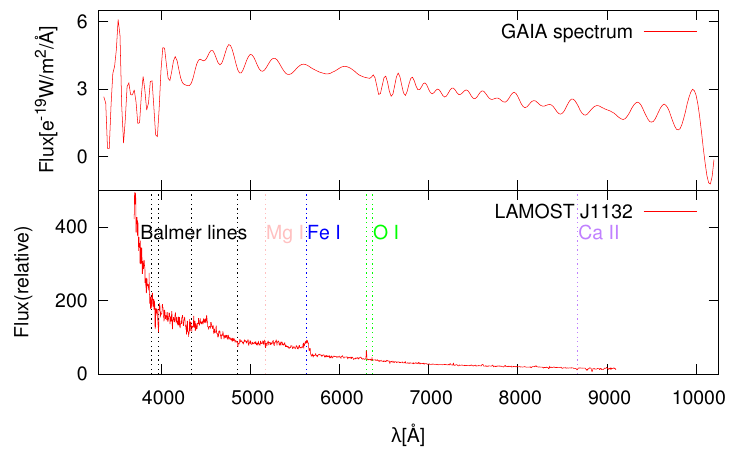}
\end{center}
\caption{LAMOST DR12 LRS (lower panel) and GAIA DR3 XP (upper panel) spectra of LAMOST J1132. The GAIA source ID for LAMOST J1132 is 3797112657191340032, with GAIA g mean magnitude of 17.26\,mag.}
\end{figure}

\begin{figure}
\begin{center}
\includegraphics[width=12cm,angle=0]{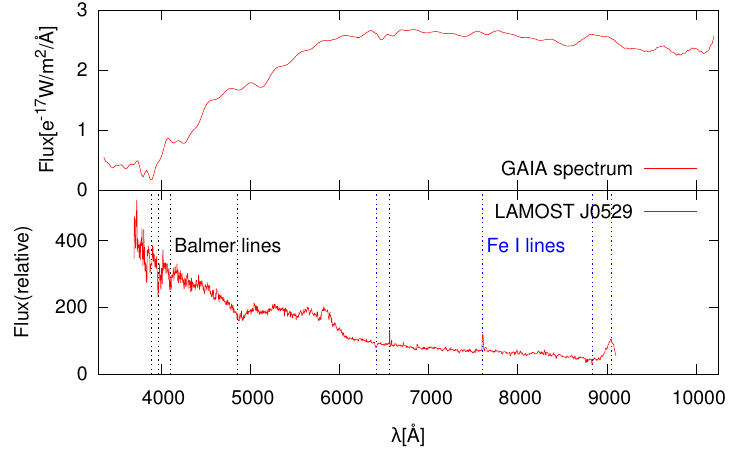}
\end{center}
\caption{LAMOST DR12 LRS (lower panel) and GAIA DR3 XP (upper panel) spectra of LAMOST J0529. The GAIA source ID for LAMOST J0529 is 182658435048551040, with GAIA g mean magnitude of 12.60\,mag.}
\end{figure}

\begin{figure}
\begin{center}
\includegraphics[width=14cm,angle=0]{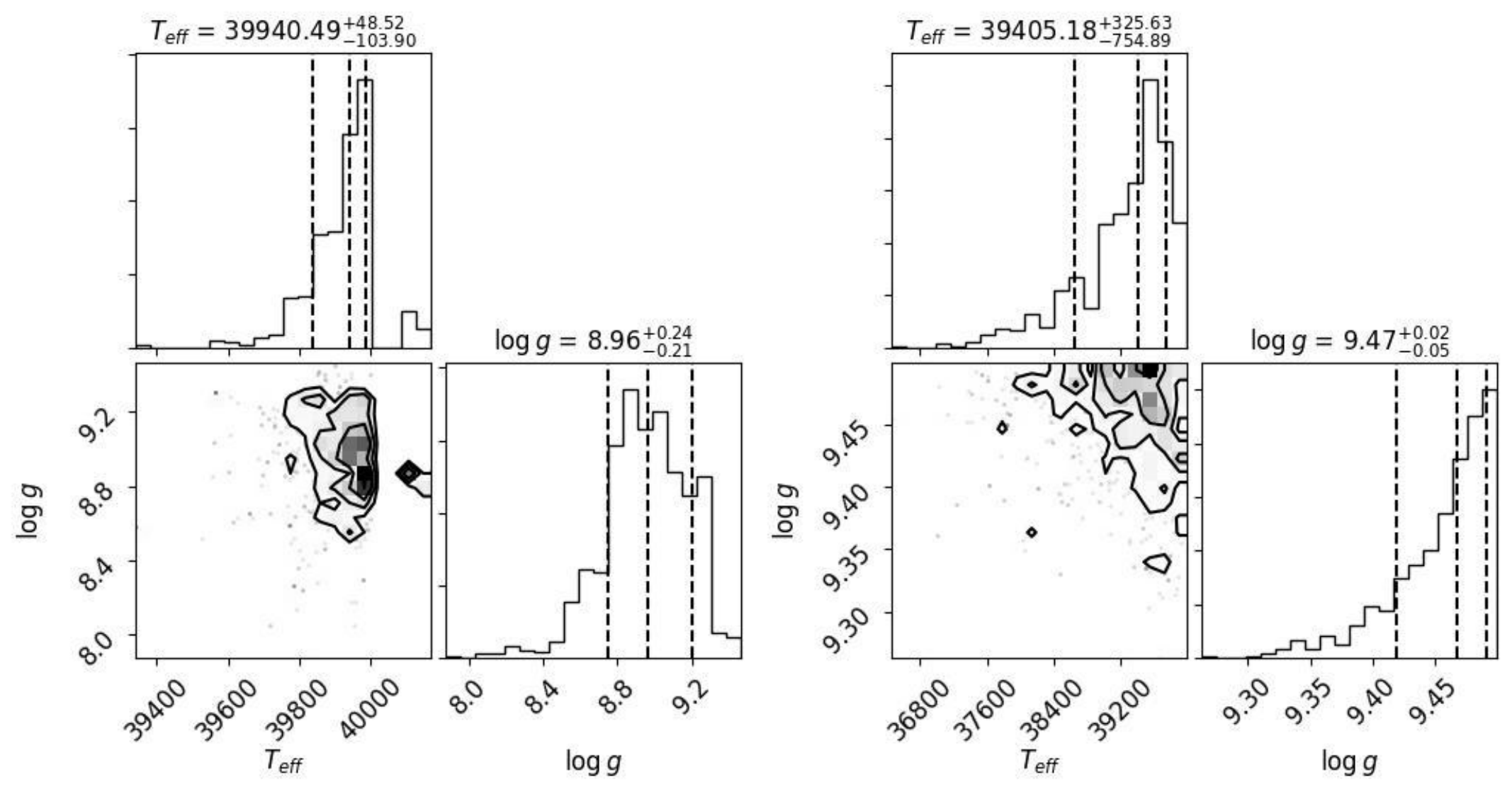}
\end{center}
\caption{The fitting corner figure for LAMOST J1132 (left) and LAMOST J0529 (right) generated by \texttt{WDTOOLS} using the GFP.}
\end{figure}

\begin{table}
\begin{center}
\caption{Absorption or emission lines identified in Fig. 1 and 2.}
\begin{tabular}{llllllllllll}
\hline
Lines                                                           &Wavelength[${\AA}$]              &Color        \\
\hline
H$\zeta$                                                        &\,3890.15                        &black        \\
H$\epsilon$                                                     &\,3971.19                        &black        \\
H$\delta$                                                       &\,4102.89                        &black        \\
H$\gamma$                                                       &\,4341.68                        &black        \\
H$\beta$                                                        &\,4862.68                        &black        \\
H$\alpha$                                                       &\,6564.61                        &black        \\
Fe I                                                            &\,5635.53                        &blue         \\
Fe I                                                            &\,6413.44                        &blue         \\
Fe I                                                            &\,7608.60                        &blue         \\
Fe I                                                            &\,8835.44                        &blue         \\
Fe I                                                            &\,9050.30                        &blue         \\
Mg I                                                            &\,5176.70                        &pink         \\
O I                                                             &\,6301.94                        &green        \\
O I                                                             &\,6365.54                        &green        \\
Ca II                                                           &\,8664.52                        &purple       \\
\hline
\end{tabular}
\end{center}
\end{table}

In Sect. 2, we plan to investigate the possible binary nature of LAMOST J1132 and LAMOST J0529 by employing methods such as spectral fitting, spectral line identification, and spectral comparative analysis.

LAMOST is ingeniously designed, enabling both a large effective aperture of 3.6-4.9 meters and a wide field of view of 5-degree (Cui et al. 2012). It can simultaneously acquire 4,000 spectra in a single exposure at R=1800, down to a limiting magnitude of r=19 (Zhao et al. 2012). On the official website of LAMOST DR12 (https://www.lamost.org/dr12/), there are 12,605,485 LRS spectra from October 24, 2011 Pilot survey to June 3, 2024 Twelfth year survey, 3,361,604 MRS non time-domain spectra, and 12,110,344 MRS time-domain spectra. This constitutes a vast and precious collection of open-source astronomical spectral data. For the LAMOST DR12 LRS spectra, the official website provides detailed spectral classification for stars (including types A, F, G, K, and M), galaxies (including galaxy population synthesis), QSOs (including QSO emission line features), WDs, and CVs, and so on. While examining high signal-to-noise ratio (S/N) magnetic WD spectra, we noticed LAMOST J1132 and LAMOST J0529. Being uncataloged in the SIMBAD astronomical database and being potential binaries based on our preliminary spectral analysis, they are compelling targets for further investigation.

In Fig. 1 and 2, we show the LAMOST DR12 LRS (lower panel) and GAIA DR3 XP (upper panel) spectra of LAMOST J1132 (GAIA source ID 3797112657191340032) and LAMOST J0529 (GAIA source ID 182658435048551040), respectively. On the official website, the LAMOST data processing and release team identified them as magnetic WDs through cross-matching large datasets. However, based on detailed spectral analysis, medium resolution or high resolution spectra may be required to confirm the magnetic field strength. The significantly elevated flux at the blue end of the continuum readily suggests that they exhibit the spectral characteristics of a hot WD. Reindl et al. (2023) reported 68 new hot WDs, and for further information on hot WDs, we can refer to this literature. Chandra et al. (2020) made a WD spectral fitting tool named \texttt{WDTOOLS} publicly available. This Python software package employs a generative fitting pipeline (GFP) and random forest regression model to determine the optimal fitting parameters for the target WD's spectrum. The GFP interpolates theoretical spectra (atmospheric models, Koester 2010) with artificial neural networks using Markov Chain Monte Carlo algorithm. The random forest regression model forms a regression relation between H Balmer lines on the spectrum and stellar labels derived by Tremblay et al. (2019). The method calculates prediction uncertainties by constructing an ensemble of 100 random forest models trained on bootstrap resamples of the training data, and then uses the variance of their predictions to define 1$\sigma$ error intervals, following the approach of Coulston et al. (2016). According to the usage examples on the official website of \texttt{WDTOOLS}, we fitted LAMOST J1132 and LAMOST J0529 with the GFP and the random forest regression model. With the GFP, we show the fitting corner figure for LAMOST J1132 (left) and LAMOST J0529 (right) in Fig. 3. Since the effective temperature ($T_{eff}$) range of the synthetic spectra for 1,274 pure-H atmospheres is 6,000-40,000\,K (Chandra et al. 2020), the fitting results in Fig. 3 suggest that two sources have reached the upper $T_{eff}$ limit. Therefore, the WDs in these two sources have $T_{eff}$ exceeding 40,000\,K. The random forest regression model fits the first four Balmer lines (H$\alpha$, H$\beta$, H$\gamma$, H$\delta$) by performing a $\chi^{2}$ minimization of symmetric Voigt profiles to each line using the PYTHON's LMFIT package (Chandra et al. 2020). The $T_{eff}$ and surface gravities (log$g$) for LAMOST J1132 and LAMOST J0529 are 53728$\pm$2467\,K and 47381$\pm$494\,K, 7.98$\pm$0.08 and 7.84$\pm$0.05, respectively. It is worth noting that the quality of the Balmer lines in the spectra of LAMOST J1132 and LAMOST J0529 is relatively low (possibly due to magnetic fields), but it can be confirmed that both sources contain a hot WD.

We have identified the absorption or emission lines in the LAMOST spectra from Fig. 1 and 2, as shown in Table 1. The corresponding wavelengths are from the LAMOST official website or the NIST Atomic Spectra Database (Kramida 2008). The presence of these neutral metal lines suggests a cool companion star, while the Ca II lines may indicate stellar activity. We retrieved the XP spectra of these two sources by their GAIA source IDs, as shown in the upper panels of Fig 1 and Fig 2. The parallaxes of these two sources measured by GAIA are 0.073$\pm$0.108\,mas and 0.632$\pm$0.017\,mas, respectively. The failure of the GAIA XP spectral shape to match that of the LAMOST spectrum also indicates that the XP spectrum is likely affected by the companion star. Therefore, based on the spectral analysis, these two sources should be binaries containing a hot WD and a cool companion star. We look forward to more observations of these two sources by LAMOST in the future, particularly MRS observations, which will help reveal more detailed physical information, such as the magnetic field properties of hot WDs. Spectral analysis is significantly important for studying astrophysics.

\section{A supplementary verification for LAMOST J1132 and LAMOST J0529 based on photometric data}

\begin{table*}
\begin{center}
\tiny
\caption{Apparent magnitudes for LAMOST J1132 and LAMOST J0529. The optical data is from SkyMapper for LAMOST J1132 and from Pan-STRRS for LAMOST J0529. The NIR data and MIR data are from 2MASS and WISE respectively for both LAMOST J1132 and LAMOST J0529. Some color index values are calculated.}
\begin{tabular}{lcccccccccccccccccccccccccc}
\hline
LAMOST J1132       &             &            &           &            &             &            &            &           &          &          &          &           &          \\
filter             &u            &v           &g          &r           &i            &z           &J           &H          &Ks        &w1        &w2        &w3         &w4        \\
$\lambda$($nm$)    &\,355.0      &\,387.0     &\,497.0    &\,604.0     &\,771.0      &\,909.0     &\,1,250     &\,1,650    &\,2,160   &\,3,400   &\,4,600   &\,12,000   &\,22,000  \\
\hline
mag                &\,18.920     &\,18.600    &\,17.543   &\,17.245    &\,17.078     &\,17.023    &\,16.203    &\,15.717   &\,15.545  &\,15.656  &\,15.705  &\,12.591   &\,8.4     \\
err                &\,0.010      &\,0.007     &\,0.003    &\,0.005     &\,0.007      &\,0.009     &\,0.084     &\,0.053    &\,0.221   &\,0.023   &\,0.055   &           &          \\
\hline
color index        &             &            &           &            &             &            &z-J         &J-H        &H-Ks      &          &          &           &          \\
\hline
mag                &             &            &           &            &             &            &0.820       &0.486      &0.172     &          &          &           &          \\
\hline
LAMOST J0529       &             &            &           &            &             &            &            &           &          &          &          &           &          \\
filter             &             &g           &r          &i           &z            &y           &J           &H          &Ks        &w1        &w2        &w3         &w4        \\
$\lambda$($nm$)    &             &\,486.6     &\,621.5    &\,754.5     &\,867.9      &\,963.3     &\,1,250     &\,1,650    &\,2,160   &\,3,400   &\,4,600   &\,12,000   &\,22,000  \\
\hline
mag                &             &\,13.5985   &\,12.7500  &\,12.2650   &\,11.9560    &\,11.7950   &\,10.603    &\,9.980    &\,9.821   &\,9.701   &\,9.793   &\,9.646    &\,8.738   \\
err                &             &\,0.0014    &           &            &             &            &\,0.020     &\,0.021    &\,0.018   &\,0.022   &\,0.022   &\,0.052    &          \\
\hline
 color index       &             &            &           &            &             &            &z-J         &J-H        &H-Ks      &          &          &           &          \\
\hline
mag                &             &            &           &            &             &            &1.353       &0.623      &0.159     &          &          &           &          \\
\hline
\end{tabular}
\end{center}
\end{table*}

\begin{figure}
\begin{center}
\includegraphics[width=12cm,angle=0]{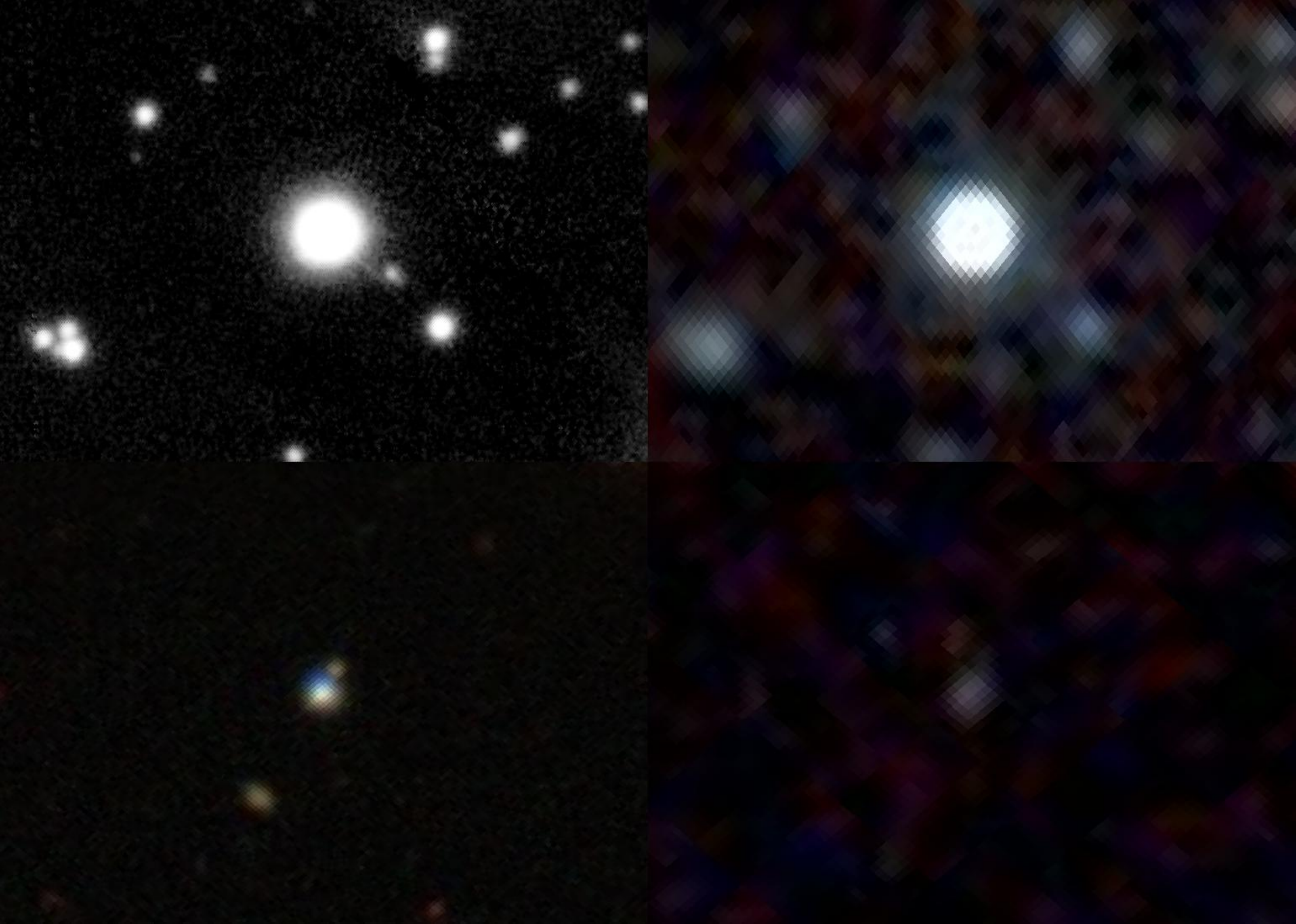}
\end{center}
\caption{Photometric observation images of LAMOST J1132 and LAMOST J0529 from the CDS portal. The two lower panels of LAMOST J1132 are from SDSS9 color (left) and 2MASS J, H, and Ks bands (right), respectively. The two upper panels of LAMOST J0529 are from Pan-STARRS DR1 g band (left) and 2MASS J, H, and Ks bands (right), respectively. Each panel covers a 1.11' field of view.}
\end{figure}

In Sect. 2, we confirmed through spectral analysis that the primary stars of both LAMOST J1132 and LAMOST J0529 are hot WDs. In Sect. 3, we plan to investigate the physical characteristics of their cool companion stars using multi-band photometric data from the optical, near-infrared (NIR), and mid-infrared (MIR) wavelengths.

Modern astronomy encompasses full-wavelength observations from gamma-ray to radio bands, and multi-wavelength studies are crucial for the comprehensive analysis of target celestial objects. For the two target sources in our study, spectroscopic information indicates the presence of a hot WD. Evidence for a cool companion star requires further confirmation. We plan to search for observational data in the optical, NIR, and MIR bands. The SkyMapper telescope (Keller et al. 2007) is designed to perform the Southern Sky Survey with six-colour (u, v, g, r, i, and z bands) and multi-epoch (four-hour, one-day, one-week, one-month, and one-year sampling) photometric survey. Developed by the University of Hawaii's Institute for Astronomy and located at the Haleakala Observatories, the Panoramic Survey Telescope and Rapid Response System (Pan-STARRS, Chambers et al. 2016) has, among its key goals, executing a precision photometric and astrometric survey of stars (g, r, i, z, and y bands) in the Milky Way and the Local Group. The Two Micron All Sky Survey (2MASS, Skrutskie et al. 2006) performed ground-based NIR survey in J, H, and Ks bands and the Wide-field Infrared Survey Explorer (WISE, Wright et al. 2010), as a spacecraft, completed a NIR survey of the entire sky in w1, w2, w3, and w3 bands.

The apparent magnitudes for LAMOST J1132 and LAMOST J0529 are shown in Table 2. LAMOST J0529 is approximately 100 times brighter than LAMOST J1132 in the optical band. It is noteworthy that the flux calibrations for a zero-magnitude star differ across the optical, NIR, and MIR bands. Both sources exhibit strong radiation in the NIR and MIR bands, indicating the presence of cool companion stars. Using a similar research methodology, we examined the dwarf nova IU Leo (Chen et al. 2026) and discovered a new binary system, LAMOST J064137.77+045743.8 (Chen et al. 2025). The apparent magnitude in the optical band is heavily influenced by hot WDs, while the spectral type of the companion star can be preliminarily determined using color indices in the NIR band. We show the z-J, J-H, and H-Ks values for LAMOST J1132 and LAMOST J0529 in Table 2. Referring to the color index table for luminosity class V stars in Covey et al. (2007), for LAMOST J1132, the values z-J=0.820 and J-H=0.486 are close to a K0-type star, while H-Ks=0.172 is close to an M0-type star. For LAMOST J0529, the values z-J=1.353, J-H=0.623, and H-Ks=0.159 are close to an M2.5-type star. Based on color index analysis, we conclude that the companion star of LAMOST J1132 is most likely of spectral type K or M, while that of LAMOST J0529 is likely of spectral type M.

Photometric observation images enable us to gain a more direct understanding of the research target and provide a more straightforward and effective means of validating our analytical process. In Fig. 4, we show the photometric observation images of LAMOST J1132 and LAMOST J0529 from the CDS portal. The two lower panels are for LAMOST J1132, from SDSS9 color (left) and 2MASS J, H, and Ks bands (right), respectively. The two upper panels are for LAMOST J0529, from Pan-STARRS DR1 g band (left) and 2MASS J, H, and Ks bands (right), respectively. LAMOST J0529 is indeed much brighter than LAMOST J1132 in Fig. 4. Both sources indeed exhibit strong radiation in the NIR bands. The strong emission in the NIR bands cannot be explained by a single hot WD. It indicates that both sources are binaries composed of a hot WD and a cool companion star.

\section{A solid evidence of the companion star for LAMOST J1132 and LAMOST J0529 based on light curve data released by ZTF telescope}

\begin{figure}
\begin{center}
\includegraphics[width=12cm,angle=0]{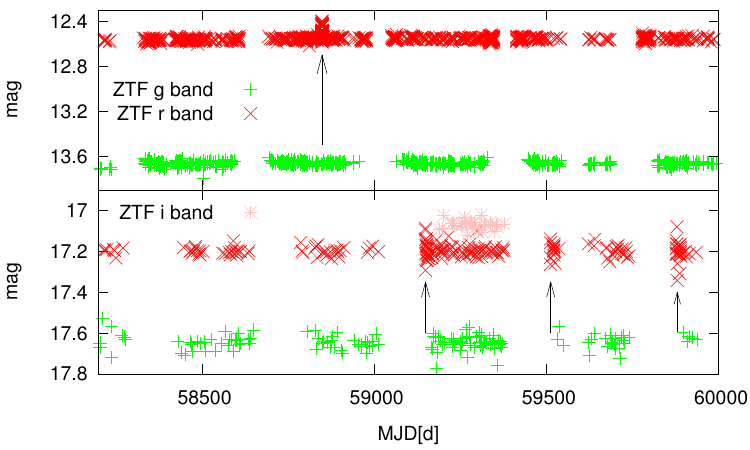}
\end{center}
\caption{Light curves of g, r, and i bands for LAMOST J1132 (lower panel) and g and r bands for LAMOST J0529 (upper panel) from ZTF telescope. The black arrows indicate the flares.}
\end{figure}

\begin{figure}
\begin{center}
\includegraphics[width=12cm,angle=0]{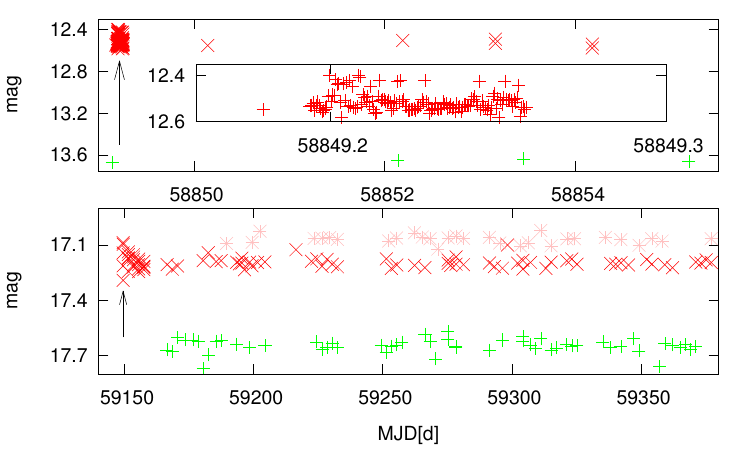}
\end{center}
\caption{A magnified view of Fig. 5, in which the upper panel further zooms in on the flare event.}
\end{figure}

\begin{figure}
\begin{center}
\includegraphics[width=12cm,angle=0]{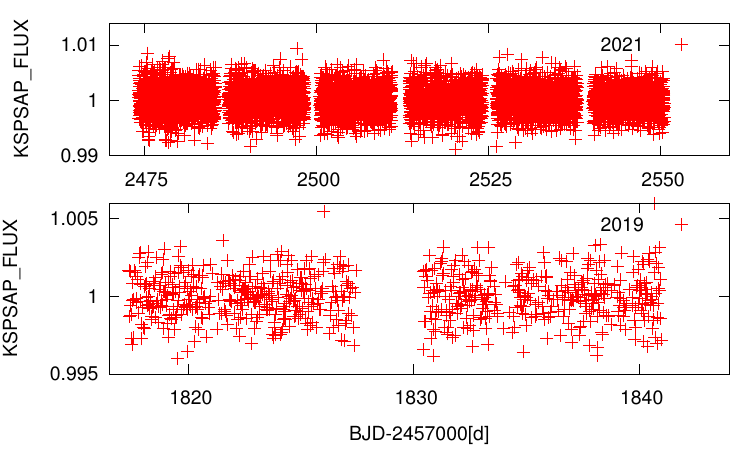}
\end{center}
\caption{Light curves of LAMOST J0529 from TESS observations in 2019 and 2021.}
\end{figure}

\begin{figure}
\begin{center}
\includegraphics[width=12cm,angle=0]{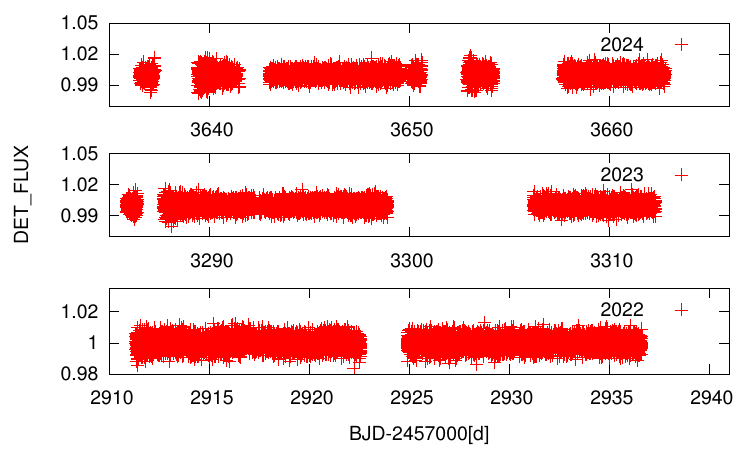}
\end{center}
\caption{Light curves of LAMOST J0529 from TESS observations in 2022, 2023, and 2024.}
\end{figure}

In Sect. 4, we aim to utilize long-term light curves of LAMOST J1132 and LAMOST J0529 to investigate physical properties such as the flaring activity of the companion stars and the orbital periods of the binary systems.

The light curve serves as the most direct evidence for a variable source. In Fig. 5, from the ZTF telescope, the lower panel displays the g, r, and i bands light curves for LAMOST J1132 and the upper panel displays the g and r bands light curves for LAMOST J0529. The process of rapid increase in brightness (with a decrease in magnitude value) in the r-band over short timescales is considered to be a stellar flare event, as indicated by the black arrows in Fig. 5. A stellar flare event occurs when magnetic field lines reconnect, which results in a large release of energy (Klein et al. 2020). Davenport (2016) reported the detection of 4,041 flaring stars in Kepler Mission (Kepler, Borucki et al. 2010) data, accounting for 1.9\% of the Kepler stellar sample. G$\ddot{u}$nther et al. (2020) identified 673 M dwarfs and 184 K type stars among 1228 flaring stars based on the first Transiting Exoplanet Survey Satellite (TESS, Ricker et al. 2014) data release. The M dwarfs flare are more frequently and strongly than other stars.

In Fig. 6, we show a magnified view of Fig. 5. For LAMOST J1132, three flares occurred at MJD=59149.5, 59512.5, and 59880, but the observational data points are sparse. These three events exhibited magnitude changes of 0.2 within 0.02 days, 0.12 within 0.01 days, and 0.25 within 1 day, respectively. The rapid, significant brightening relative to the mean magnitude on such short timescales, as shown in the lower panel of Fig. 5 and 6, makes it highly probable that they are flares, even though the limited data points cannot reveal the detailed profile of the flare light curves. For LAMOST J0529, the flare occurred at MJD=58849.2 and a magnitude change of 0.2 occurred within 0.07 days. Although the data points are relatively dense, as shown in the upper panel of Fig. 6, they still cannot outline the detailed profile of the flare like space telescopes (Kepler or TESS) can. Based on the Kepler data, Hawley et al. (2014) employed an equivalent duration method to calculate flare energy for individual stars. We can roughly estimate the rise and decay durations of the flare in the upper panel of Fig. 6. For both LAMOST J1132 and LAMOST J0529, the r-band extends to 570\,nm (Bellm et al. 2019), where emission from the hot WD is still significant, as shown in the lower panel of Fig. 1 and 2. The measured quiescent luminosity is that of the entire binary system. Hence, the flare luminosity of the individual star cannot be isolated, preventing the calculation of a specific flare energy.

For LAMOST J1132, the observed g-band magnitude variation is $\sim$0.15 mag, obviously exceeding the observational error of $\sim$0.025 mag. For LAMOST J0529, the magnitude variation in the g-band ($\sim$0.04 mag) is on the same order of magnitude as the observational error ($\sim$0.01 mag). For LAMOST J1132, we tend to believe that the optical variability in the g-band is not caused by observational errors, but rather by a combined effect of binary motion and companion star flares. Rapidly rotating stars can generate starspots and modulate stellar brightness through rotation (Davenport et al. 2016). Rapidly rotating M dwarfs are most prone to producing flares, and their flare amplitude is independent of the rotation period (G$\ddot{u}$nther et al. 2020). In Table 3, we list the period information obtained from the light curve variation provided on the ZTF webpage. Based on the observed periods, we infer that the binary orbital period of LAMOST J1132 may be on the order of days or on the order of tens of days. The 2.3\,days period in the r-band is likely the rotation period of the companion star. The time intervals between the three flares are 363 days and 367.5 days respectively, which appear to be a one-year magnetic activity cycle. The Sun has a rotation period of 25.4 days and a magnetic activity cycle of 11 years (Noyes \& Weiss 1984). Although our inferred results are on the same order of magnitude as the rotation period (0.45875 days) and magnetic activity cycle ($\sim$350 days) of the dM1-2e star EY Dra (Vida et al. 2010), the companion star's rotation period and magnetic activity cycle for LAMOST J1132 are preliminary estimates and require further confirmation. Applying the phase-folding iteration method to test each of these periods did not yield a phased light curve with a single-period signal. The light curve signal is likely a composite of the binary orbital motion, modulation from the companion star's rotation, and flares of the companion star. For LAMOST J0529, the magnitude variation is not significantly greater than the observational errors, and the obtained periods in Table 3 show no consistent period. Based on a comprehensive analysis of LAMOST J1132, for the periods associated with LAMOST J0529 listed in Table 3, the ones on the order of years to decades may be magnetic activity cycles or binary orbital periods. The period on the order of months could be the rotational period of the companion star, while the period on the order of hours might be induced by the magnetic activity of microflares. It is a complex multi-periodic coupling of light variations and these periods are preliminary inferences and require further confirmation. The lack of obvious eclipses explains why these two sources were not noted by previous studies or recorded in the SIMBAD astronomical database. According to Newtonian mechanics, the motion of binary stars follows an equation below,
\begin{equation}
\frac{G(M_{1}+M_{2})}{L^{3}}=\frac{4\pi^{2}}{T^{2}}.
\end{equation}
\noindent In Eq.\,(1), G is the gravitational constant, $M_{1}$ and $M_{2}$ are the masses of the binary stars, L is the separation between the two stars, and T is the orbital period of the binary system. Taking the mass of the hot WD as 0.6\,$M_{\bigodot}$ and the companion star's mass as 0.4\,$M_{\bigodot}$ (cool red dwarfs generally have small masses and radii, Chabrier \& Baraffe 2000), if we perform rough and preliminary calculations for orbital periods of 10\,days, 365\,days, and 1,000\,days, respectively, the distances between the binary stars are approximately 19.5, 215, and 421\,$R_{\bigodot}$. The radii of both component stars in the two binary systems are smaller than $R_{\bigodot}$ (the radius of a WD is comparable to that of Earth), while the separations between the two stars of the two binary systems are far larger than $R_{\bigodot}$. As a result, the probability of mutual eclipses is extremely low, making LAMOST J1132 and LAMOST J0529 non-eclipsing binary systems.

\begin{table}
\begin{center}
\caption{Derived periods for LAMOST J1132 from ZTF and for LAMOST J0529 from ZTF and TESS.}
\begin{tabular}{llllllllllll}
\hline
ZTF                             &LAMOST J1132                     &                      &                      &LAMOST J0529               &           \\
Filter(Duration)                &Period[d]                        &S/N                   &Filter(Duration)      &Period[d]                  &S/N        \\
\hline
g(1733.224\,days)               &\,17.23                          &\,6.293               &g1(442.828\,days)     &\,10.07                    &\,8.475    \\
                                &\,9.861                          &\,4.955               &                      &\,9.577                    &\,7.978    \\
                                &\,7.93                           &\,4.827               &g2(1787,012\,days)    &\,1630                     &\,8.868    \\
                                &\,11.35                          &\,4.395               &                      &\,326                      &\,8.395    \\
                                &\,13.4                           &\,4.288               &g3(1634.731\,days)    &\,381.3                    &\,15.86    \\
r(1718.187\,days)               &\,2.333                          &\,5.999               &                      &\,126.1                    &\,11.49    \\
                                &\,2.233                          &\,5.873               &r1(449.832\,days)     &\,30.19                    &\,5.122    \\
                                &\,2.378                          &\,5.831               &                      &\,7.048                    &\,4.696    \\
                                &\,2.325                          &\,5.549               &r2(1752.987\,days)    &\,924.9                    &\,18.16    \\
                                &\,2.218                          &\,5.487               &                      &\,2929                     &\,14.49    \\
i(187.665\,days)                &\,15.53                          &\,3.383               &r3(1768.069\,days)    &\,1.003                    &\,8.716    \\
                                &\,10.73                          &\,3.252               &                      &\,0.2496                   &\,8.523    \\
                                &\,26.59                          &\,3.057               &                      &                           &           \\
                                &\,11.69                          &\,2.911               &                      &                           &           \\
                                &\,8.865                          &\,2.761               &                      &                           &           \\
\hline
TESS                            &                                 &                      &                      &LAMOST J0529               &           \\
                                &                                 &                      &year                  &Period[d](Amplitude[ppt])  &S/N        \\
\hline
                                &                                 &                      &2019                  &0.485(0.277)               &2.961      \\
                                &                                 &                      &                      &0.390(0.283)               &3.045      \\
                                &                                 &                      &2021                  &0.120(0.269)               &6.427      \\
                                &                                 &                      &                      &0.817(0.235)               &5.614      \\
                                &                                 &                      &2022                  &0.886(0.296)               &4.495      \\
                                &                                 &                      &                      &0.120(0.283)               &4.336      \\
                                &                                 &                      &2023                  &0.005(0.317)               &3.848      \\
                                &                                 &                      &                      &0.657(0.315)               &3.678      \\
                                &                                 &                      &2024                  &0.408(0.377)               &3.949      \\
                                &                                 &                      &                      &0.342(0.401)               &4.212      \\
\hline
\end{tabular}
\end{center}
\end{table}

The Barbara A. Mikulski Archive for Space Telescopes (MAST, Hou et al. 2023) archives observational data from numerous telescopes, including the Hubble Space Telescope (HST, Shore 1992), the Kepler mission, and TESS, and so on, and provides user-friendly access. We attempted to search for basic information on these two sources in MAST and found no records for LAMOST J1132, while there were seven TESS observation data entries for LAMOST J0529, with one observation in 2019, three in 2021, and one each in 2022, 2023, and 2024. The flare shown in the upper panel of Fig. 5 occurred on January 1, 2020, while the observation times of these seven TESS datasets do not coincide with it. We selected observational data with quality=0 and plotted the light curves using KSPSAP\_Flux for 2019 (24\,days) and 2021 (76\,days) data, and DET\_Flux for 2022 (25\,days), 2023 (26\,days), and 2024 (26\,days) data, as shown in Fig. 7 and Fig. 8 respectively. The high-precision, long-baseline light curve from TESS for LAMOST J0529, like the ZTF data, show no apparent eclipsing signals. Period04 (Lenz \& Breger 2005) is a software package designed for sophisticated time string analysis. We utilized the Period04 program to extract frequency signals from the time-domain data of LAMOST J0529 observed by TESS. The two periods with the highest S/N together with corresponding amplitude extracted from the observational data of each year are listed in Table 3. We set the box size to 200 to extract frequencies with Period04. For data from ground-based telescopes, frequencies with a S/N greater than $\sim$4.0 (Lenz \& Breger 2005) are typically adopted as eigenfrequencies. For data from space telescopes such as TESS, frequencies with a S/N exceeding 4.6-5.7 (Baran \& Keon 2021) are generally taken as eigenfrequencies. Only the 76-day observation data from 2021 yielded eigenfrequencies with a reliable S/N. However, the 0.120-day period is too short to be interpreted as the companion star rotation period, and certainly not as the binary orbital period. For LAMOST J0529, there is no indication of matching periods extracted from the ZTF and TESS observational data, which are likely significantly influenced by stellar magnetic activity. The magnetic activity of microflares may be substantial, as shown in the magnified view of the upper panel of Fig. 6.

\section{An analysis, a discussion, and conclusions}

\subsection{A thorough analysis}

\begin{table}
\tiny
\begin{center}
\caption{Spectral and light curve data statistics table for LAMOST J1132 and LAMOST J0529.}
\begin{tabular}{llllllllllll}
\hline
Telescopes                &Observed times                                    &Instruments                       &Bands                                            &Exposure times                        \\
\hline
LAMOST J1132              &                                                  &                                  &                                                 &                                      \\
\hline
LAMOST                    &2024-05-02                                        &16 low resolution spectrographs   &u,g,r,i,z (S/N=5.50, 8.87, 13.15, 16.94, 9.38)   &5400\,s                               \\
GAIA                      &synthesize multiple observations                  &Blue photometer and red photometer&330-680\,nm for BP and 640-1050\,nm for RP       &cumulative integration time           \\
ZTF                       &2018-03-27\,(lasting 1733\,days)                  &ZTF-g filter                      &roughly 400-550\,nm                              &30\,s                                 \\
ZTF                       &2018-04-09\,(lasting 1718\,days)                  &ZTF-r filter                      &roughly 560-720\,nm                              &30\,s                                 \\
ZTF                       &2020-12-06\,(lasting 188\,days)                   &ZTF-i filter                      &roughly 700-850\,nm                              &30\,s                                 \\
\hline
LAMOST J0529              &                                                  &                                  &                                                 &                                      \\
\hline
LAMOST                    &2023-10-18                                        &16 low resolution spectrographs   &u,g,r,i,z (S/N=5.29, 9.56, 16.14, 16.98, 11.87)  &1800\,s                               \\
GAIA                      &synthesize multiple observations                  &Blue photometer and red photometer&330-680\,nm for BP and 640-1050\,nm for RP       &cumulative integration time           \\
ZTF                       &2018-03-29\,(lasting 1787\,days)                  &ZTF-g filter                      &roughly 400-550\,nm                              &30\,s                                 \\
ZTF                       &2018-04-08\,(lasting 1768\,days)                  &ZTF-r filter                      &roughly 560-720\,nm                              &30\,s                                 \\
TESS                      &2019-11-28\,(lasting 24\,days)                    &Photometer                        &600-1000\,nm                                     &1800\,s                               \\
TESS                      &2021-09-16\,(lasting 76\,days)                    &Photometer                        &600-1000\,nm                                     &600\,s                                \\
TESS                      &2022-11-26\,(lasting 25\,days)                    &Photometer                        &600-1000\,nm                                     &200\,s                                \\
TESS                      &2023-12-07\,(lasting 26\,days)                    &Photometer                        &600-1000\,nm                                     &200\,s                                \\
TESS                      &2024-11-21\,(lasting 26\,days)                    &Photometer                        &600-1000\,nm                                     &200\,s                                \\
\hline
\end{tabular}
\end{center}
\end{table}

\begin{figure}
\begin{center}
\includegraphics[width=12cm,angle=0]{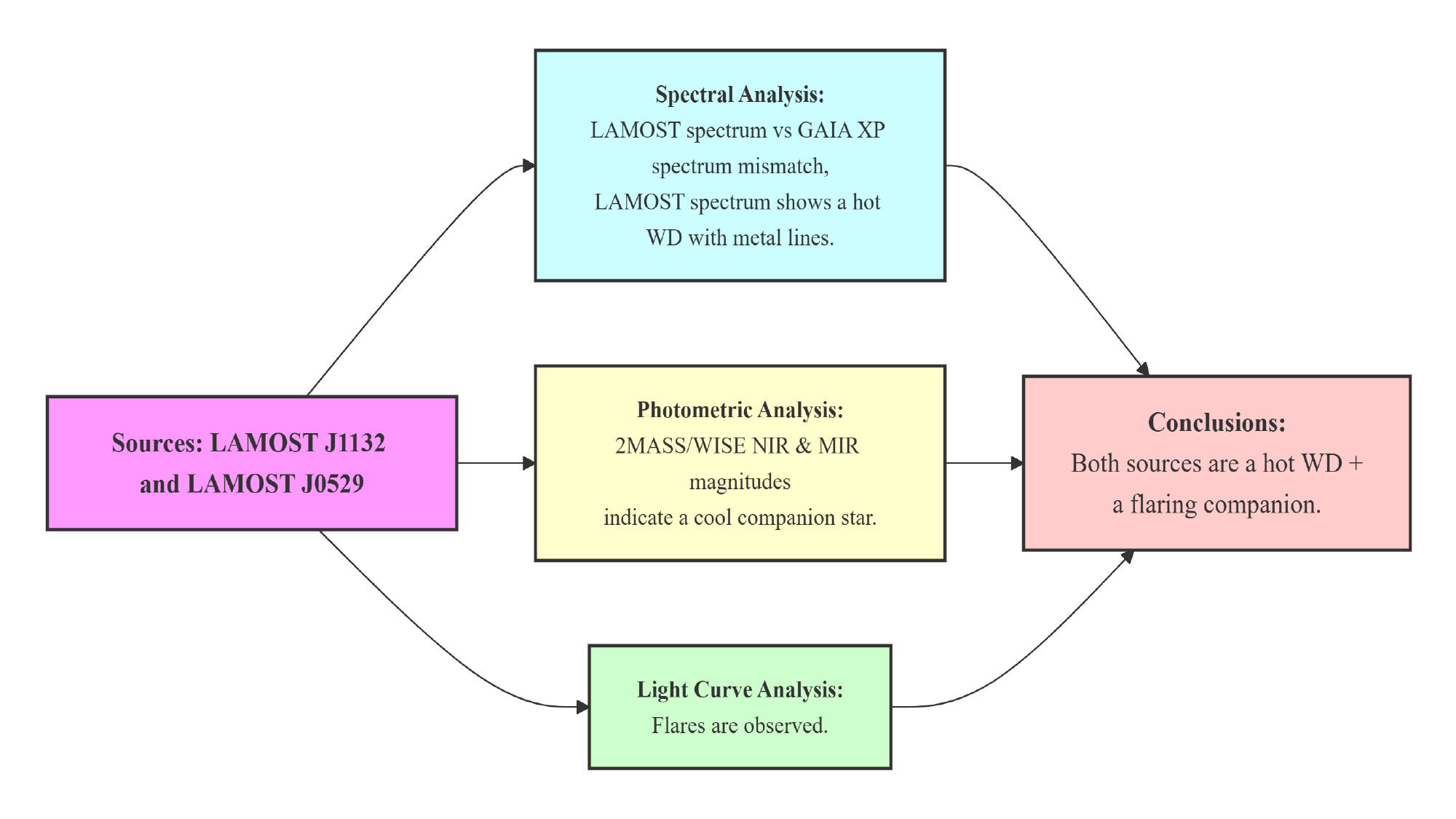}
\end{center}
\caption{Detailed analysis flowchart.}
\end{figure}

With the continuous public release of observational data from various survey telescopes, astronomical research has entered the era of big data for the broader community of astronomers. Multi-method investigations, such as spectral analysis, photometric analysis, and light curve analysis, combined with multi-wavelength studies have become more powerful approaches to uncovering the laws of astrophysics. LAMOST J1132 and LAMOST J0529 have not yet been recorded in the SIMBAD astronomical database. By utilizing spectral analysis, photometric analysis, and light curve analysis methods, we conducted a comprehensive study of these two sources and determined that they are newly discovered binary systems, each consisting of a hot WD and a companion star exhibiting flare activity.

The spectral and light curve data used are summarized in Table 4. The optical, NIR, and MIR photometric magnitude data for LAMOST J1132 were obtained from the SkyMapper telescope website, while those for LAMOST J0529 were sourced from the VizieR website. A detailed analysis flowchart is presented in Fig. 9. For spectral analysis, the inconsistency between the GAIA XP spectra and the LAMOST spectra indicates the presence of a companion star and the LAMOST spectrum of the hot WD, which exhibits metal lines, also indicates the presence of a binary system. For photometric analysis, both the NIR photometric images (Fig. 4) from 2MASS and the photometric magnitude data in the NIR (2MASS) and MIR (WISE) bands suggest that these two sources indeed harbor a cool companion system. For light curve analysis, the discovery of flare activity provides further evidence for the existence of a cool companion star. The multi-method investigations collectively demonstrate that both LAMOST J1132 and LAMOST J0529 are newly discovered binary systems.

\subsection{A concise discussion}

The spectral fitting results of two sources using the \texttt{WDTOOLS} program, based on the LAMOST spectra, indicate that both WDs have $T_{eff}$ exceeding 40,000\,K, likely around 53728$\pm$2467\,K for LAMOST J1132 and 47381$\pm$494\,K for LAMOST J0529. In the optical band, the WD dominates the flux. By analyzing the apparent magnitude differences in the NIR band, it can be estimated that the spectral type of the companion star of LAMOST J1132 is K or M-type, while that of LAMOST J0529 is M-type. We have preliminarily identified 3 stellar flares for LAMOST J1132 and 1 stellar flare for LAMOST J0529. The flare phenomenon is very common in M-type red dwarfs, which is self-consistent with the identified spectral types of the companion stars from the apparent magnitude differences.

As reported by El-Badry \& Rix (2018), MS-MS binaries are the most numerous among binary systems, while WD-MS binaries constitute a significant component of such systems. Based on multi-wavelength observations, astrometry, and photometry, Nayak et al. (2024) investigated 93 WD-MS binary candidates, finding that most of the MS companions are of spectral types K or M. For WD-MS binaries, research has primarily focused on the physics of accretion (Webb 2023), specifically within the realm of CVs. Based on GAIA DR3 astrometry, SDSS-V spectroscopy, and LAMOST DR6 spectroscopy, 1,587 CVs (Canbay et al. 2023), 504 CVs (Inight et al. 2025), and 323 CV candidates (Sun et al. 2021) have been intensively studied, respectively. Research on systems without accretion or orbital eclipses, such as LAMOST J1132 and LAMOST J0529, remains quite limited. This is of great significance for studying magnetic activity in binary star systems and uncovering more comprehensive physical principles governing such systems.

\subsection{Clear conclusions}

Large-sample spectroscopic studies alone can easily misidentify sources like LAMOST J1132 and LAMOST J0529 as hot WDs, while large-sample photometric studies focusing on light curves are prone to misclassifying them as single stars due to the absence of observed eclipses. Comprehensive analysis reveals that these two sources are actually binary systems consisting of a hot WD and a flaring companion star. The companion star in LAMOST J1132 is most likely K or M-type, while that in LAMOST J0529 is most likely to be M-type. Applying the phase-folding iteration method did not yield a single periodic signal. Their light curves exhibit mixed signals originating from the orbital period, the rotation period of the cool companion star (which may modulate the light curve), and the magnetic activity cycle of the cool companion. Calculations based on the order-of-magnitude estimates of their orbital periods indicate an extremely low eclipse probability, confirming that both LAMOST J1132 and LAMOST J0529 are non-eclipsing binary systems.

We are sincerely grateful for the open-source data and software, as these two sources represent a newly discovered binary system containing a hot WD and a flaring companion star. The continuous release of various open-access big data, such as LAMOST MRS data, will facilitate the study of magnetic field information and time-resolved spectroscopy.

\section{Acknowledgment}

Guoshoujing Telescope (the Large Sky Area Multi-Object Fiber Spectroscopic Telescope LAMOST) is a National Major Scientific Project built by the Chinese Academy of Sciences. Funding for the project has been provided by the National Development and Reform Commission. LAMOST is operated and managed by the National Astronomical Observatories, Chinese Academy of Sciences. This work is supported by the International Centre of Supernovae, Yunnan Key Laboratory (No. 202302AN36000101) and the Yunnan Provincial Department of Education Science Research Fund Project (No. 2024J0964).

\label{lastpage}

\end{document}